# Text2VP: Generative AI for Visual Programming and Parametric Modeling

Guangxi Feng[1], Wei Yan[2]

[1] Department of Architecture, Texas A&M University, 400 Bizzell St, College Station, Texas, feng.g@tamu.edu

[2] Department of Architecture, Texas A&M University, 400 Bizzell St, College Station, Texas, wyan@tamu.edu

**Abstract**

The integration of generative artificial intelligence (AI) into architectural design has witnessed a significant evolution, marked by the recent advancements in AI to generate text, images, and 3D models. However, no prior AI applications exist for text-to-parametric models that are used in architectural design for generating various parametric design options, including free-form designs, and optimizing the design options. This study creates and investigates an innovative application of generative AI in parametric modeling by leveraging a customized Text-to-Visual Programming (Text2VP) GPT derived from GPT-4.1-2025-04-14. The primary focus is on automating the generation of graph-based visual programming workflows, including parameters and the links among the parameters, through AI-generated scripts, accurately reflecting users' design intentions and allowing the users to change the parameter values interactively. The Text2VP GPT customization process utilizes detailed and complete documentation of the visual programming language components, specific instructions, and example-driven few-shot learning. Our testing demonstrates Text2VP's capability to generate working parametric models. The paper also discusses the limitations of Text2VP; for example, more complex parametric model generation introduces higher error rates. This research highlights the potential of generative AI in visual programming and parametric modeling and sets a foundation for future enhancements to handle more sophisticated and intricate modeling tasks effectively. The study aims to allow designers to create and change parametric design models without significant effort in learning a specific programming platform, such as Grasshopper.

## 1. Introduction

As reviewed (Castro Pena et al. 2021), Gero completed the first published approach to applying artificial intelligence (AI) in architectural design in 1995, utilizing genetic algorithms to generate floor plans (Gero and Schnier 1995). With the recent surge in generative artificial intelligence, notably marked by the public release of Generative Pre-trained Transformer 3.5 (GPT-3.5) in March 2022, the broader application of generative AI has been catalyzed across many aspects of the architecture, engineering, and construction (AEC) industry (Castro Pena et al. 2021; Rudolph, Tan, and Tan 2023; Saka et al. 2023). One of the most recognized advantages of generative AI is its capability to assist in computer programming with languages such as Python and C# to achieve desired functions (Ebert and Louridas 2023; Russo 2024).

In architectural design, although programming languages have not been widely and directly applied in practice, visual programming tools like Grasshopper and Dynamo have gained popularity due to their visually intuitive nature (Dong and Agogino 1997; Asl et al. 2015; Wortmann 2018; Tagashira, Kawashima, and Yasuda 2019). These tools are especially powerful in parametric modeling. However, the steep learning curve of these visual programming tools limits their broader application despite their significant advantages in design, modeling, simulation, and optimization.

This study explores an innovative approach to the application of generative AI in architectural design. It leverages generative AI's capabilities in coding scripts (such as C#) to create modeling workflows in a widely used visual programming and parametric modeling tool - Grasshopper (Rutten 2024) - that accurately and parametrically represent the user's design and modeling intentions.

To achieve this, the research has produced Text-to-Visual Programming (Text2VP) by customizing the Text2VP GPT model from the standard GPT-4.1 model (OpenAI 2025) to understand the user's modeling intentions, and to code the C# scripts to generate a Grasshopper workflow, which further generates 2D or 3D geometric models in NURBS modeling tool Rhinoceros (Robert McNeel & Associates (TLM, Inc.) 2023).

## 2. Methodology

The research is structured into two primary sections: customization of the GPT model specifically for Text2VP, and subsequent testing of its performance. The customization phase involves preparing and utilizing two distinct types of materials:

**Knowledge Base:** This serves as a detailed information resource, enabling the GPT model to retrieve essential data. Specifically, this includes:

- A detailed manual documenting all Grasshopper components in vector stores.

**Instructions for Customizing GPT:** These provide explicit, high-level guidance on the expected model behavior, including:

- Contextual data outlining the system rules, which define the scope and methodology for performing Text2VP tasks. This context describes precisely how the model should access and apply information from the knowledge base, along with illustrative few-shot learning examples.
- Four carefully curated few-shot learning examples. Each example clearly pairs user-input modeling intentions with the desired corresponding output—a generated C# script—allowing the model to recognize and internalize patterns for application in subsequent prompts.

Together, these materials underpin the customization process, effectively guiding the pretrained standard GPT model towards the specialized task of Text2VP. The ultimate goal of this tailored approach is to ensure accurate and effective generation of C# scripts for parametric modeling within the Grasshopper environment.

## 2.1. Knowledge Base

Preliminary tests indicate that the standard GPT model has a basic understanding of parametric modeling and visual programming in Grasshopper, sufficient to assist with general questions. However, to accomplish the tasks of this research—specifically coding C# scripts to generate a complete, user-intentioned parametric modeling workflow automatically—Text2VP requires more detailed information on each Grasshopper component. In visual programming, components serve as the basic units of a workflow, enabling functions such as modeling, transforming, and data processing. By connecting the inputs and outputs of components in a certain order on the canvas, a workflow for specific parametric modeling functions is achieved. The lack of complete knowledge of Grasshopper components in the pre-trained GPT model is a challenge in this research.

At the early stage of the research, no comprehensive documentation of all Grasshopper components is publicly available. Consequently, we coded a C# script to automatically traverse all Grasshopper components and retrieve and compile all necessary information about each Grasshopper component from its settings, descriptions, and help documents. This script iterates through all 1,139 originally built-in Grasshopper components, summarizing and organizing them into a single document in a unified format to customize the Text2VP GPT model with the domain-specific, foundational knowledge of visual programming in Grasshopper. The collected Grasshopper information is divided into metadata and parameter information (Figure 1). Additionally, this structured information was stored in vector stores to enable Text2VP to efficiently retrieve("Retrieval - OpenAI API" 2025) relevant Grasshopper component data from the knowledge base through semantic search.

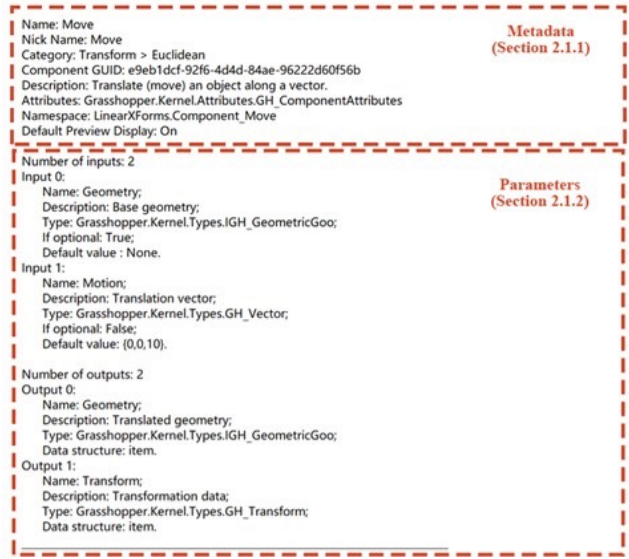

Figure 1.   An Example of Grasshopper Component Information: the "Move" Component.

### 2.1.1. Metadata

The metadata of each component includes basic information such as "Name", "Nickname", and "Category", which is usually directly displayed on the Grasshopper canvas. This bridges Text2VP to connect detailed information with its pre-trained knowledge of Grasshopper visual programming. Following this is the "Description," which briefly outlines the function of each component, aiding Text2VP in understanding the component's function. The subsequent lines reveal essential C# coding knowledge in the Grasshopper environment, specifically "Attributes" and "Namespace". Correctly using the namespace is crucial to activating the intended Grasshopper components. The final line of general information, "Default Preview Display", indicates the default display status of the components. Displaying certain modeling results, such as a clear and accurate output as user intended, is important in a visual programming parametric workflow. This knowledge enables Text2VP to control the display mode of each Grasshopper component efficiently and accurately.

### 2.1.2. Parameters

Understanding the parameters of each Grasshopper component is crucial for connecting them to form a complete parametric modeling workflow. The research gathers detailed information on each component's inputs and outputs. For both inputs and outputs, basic information includes the index for accurately referencing specific parameters; "Name" and "Description", providing an overview of each parameter; and "Type", indicating the designated data type for the parameter, as Grasshopper components only function correctly when inputs are of the proper type. For inputs, additional information includes "If optional," indicating the necessity of each input for the component to run properly. This knowledge is needed for Text2VP to ensure all necessary inputs are provided to have the Grasshopper components properly functioning. The "Default value" reveals the default setting for each input parameter, which can be used without further action if it meets the requirements, enabling Text2VP to compose the workflow in an efficient manner. For outputs, information on "Data structure" is specifically included, as variations in data structures can significantly impact the function of Grasshopper components.

### 2.2. Instructions for Customizing GPT

The second category of material used to customize the Text2VP GPT model is a set of specifically drafted descriptive instructions. These instructions clearly define the goals and tasks of Text2VP, as well as the methodological approach to be followed. They also elaborate on the structure and use of the knowledge base and the integration of few-shot learning examples. Incorporated as the "instructions" parameter in the OpenAI API, this document provides the model with high-level behavioral guidance during response generation. From a prompt engineering perspective, the instructions were carefully structured into the following distinct sections and formatted in Markdown to maximize clarity and enhance the model's ability to interpret and apply the intended guidance.

#### 2.2.1. Identity

The first section in the instructions for customizing GPT describes the personality of Text2VP for this research. It outlines the overarching goal of Text2VP as assisting users in fulfilling parametric modeling needs by composing a Grasshopper visual programming workflow through C# script. This section also briefly summarizes the specific information and knowledge that Text2VP should use, in addition to its pre-trained knowledge, to complete the specific task better. It defines the specific role of Text2VP and provides clues about the information that should be used to establish this role.

#### 2.2.2. Instructions

The subsequent section provides explicit guidance for Text2VP on generating the final C# scripts that automate the creation of Grasshopper workflows to meet user's specific parametric modeling intentions (Figure 2). This section clearly delineates the rules the model should follow, specifying both the actions it must perform and those it should explicitly avoid.

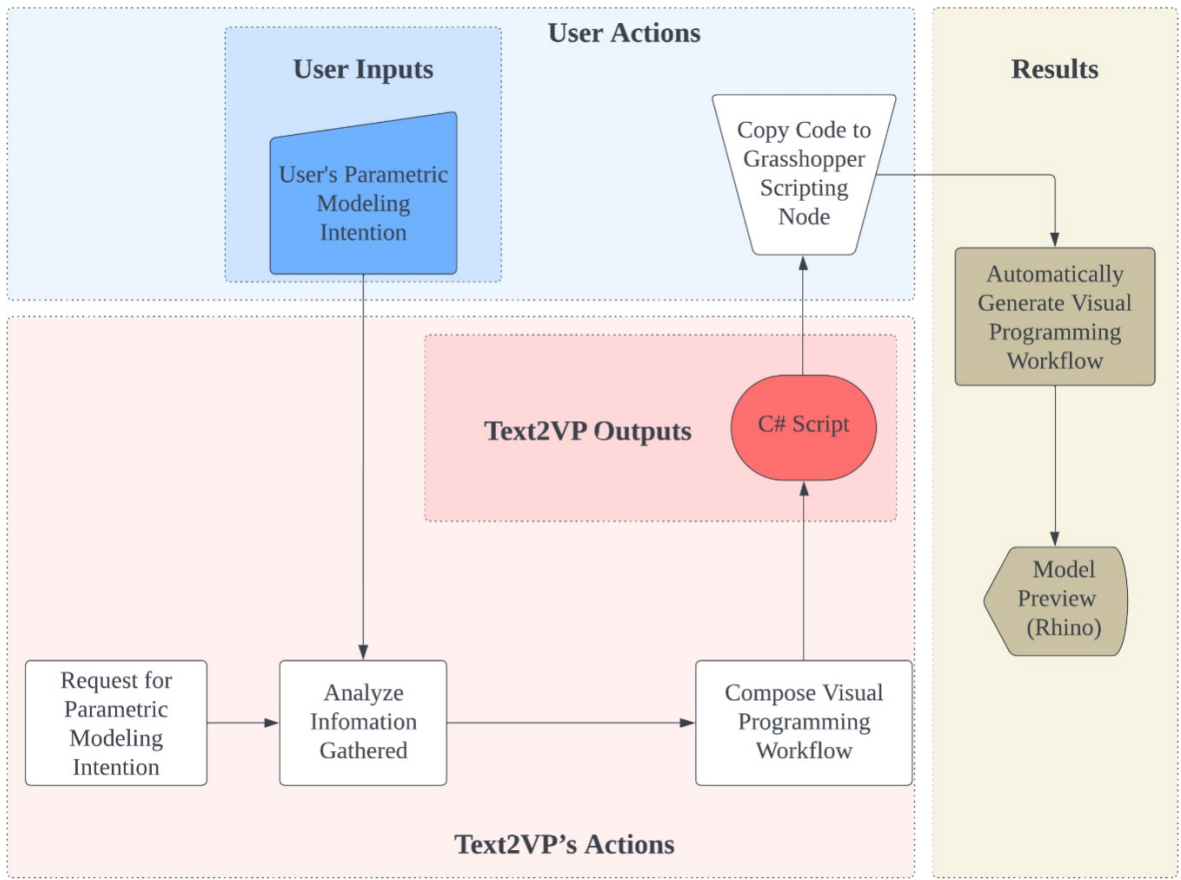

Figure 2.   Diagram of Text2VP's Tasks in Steps in Assisting Parametric Modeling in the Grasshopper Visual Programming Environment.

**Workflow and Output Requirements:** This section outlines Text2VP's primary task: composing a Grasshopper visual programming workflow and generating the corresponding C# scripts. The model is instructed to respond with scripts that users can directly implement to create parametric modeling workflows on the Grasshopper canvas.

**Reference Provided File Usage:** This section offers detailed guidance on how the Text2VP model should generate responses using information obtained through semantic search from the vector stores containing the documentation of Grasshopper components. The model is explicitly constrained to rely solely on the information from the provided file as its exclusive source of knowledge of Grasshopper components, rather than relying on pre-trained but potentially inaccurate knowledge. Furthermore, this section highlights the importance of correctly handling Grasshopper component namespaces, emphasizing that improper use of namespaces was a significant source of error observed in this research—a topic that will be addressed in detail later.

**Component Creation and Behavior, Scripting Standards, Canvas Layout & Display Rules, and Reusability and Efficiency**: These four subsections together form the core of the instructions for customizing GPT. They provide comprehensive documentation and explanations covering essential C# syntax for use in Grasshopper, including illustrative examples and guidance for special cases—such as activating and connecting Grasshopper components via namespaces, positioning components on the canvas using location coordinates that reflect their logical function within the visual programming workflow, and managing the display status of components to ensure that necessary modeling results are clearly presented while minimizing disruptions from intermediate steps. This section also explicitly enumerates forbidden actions, developed through iterative testing and analysis, to suppress undesirable outputs inherited from the model's pre-trained knowledge. For example, the use of "Rhino functions" in C# code is strictly prohibited, as this research is focused exclusively on parametric modeling within the Grasshopper environment. Overall, all content in this section is designed to thoroughly and precisely customize the Text2VP GPT model, ensuring it can most effectively assist users by automatically generating Grasshopper visual programming workflows through AI-generated C# scripts.

**Few-Shot Learning via Examples:** This last section details how Text2VP should utilize the input/output examples provided in the prompt to identify and learn patterns. The Text2VP model is instructed to draw from these examples to effectively generate Grasshopper workflows using C# scripts, applying recognized patterns to new user prompts.

### 2.2.3. Examples

As demonstrated (Brown et al. 2020), GPT achieves robust performance in a few-shot setting by significantly improving the efficiency of large AI models using in-context information, thereby enhancing the accuracy of generated results. This research adopts the in-context approach to efficiently customize the GPT model for the specific task of assisting in generating parametric modeling workflows in the visual programming environment. This is done by preparing four examples of parametric modeling needs and the corresponding C# scripts that can generate Grasshopper workflows to fulfill these needs.

The four examples for few-shot learning in this research focus on 2D and 3D parametric modeling with increased complexity. In preparing these examples, we ensure the inclusion of Grasshopper components from all fundamental function categories. These categories include parameter and input components like "Number Slider," geometry drawing components like "Polygon," geometry transformation components like "Move," algebraic data generating components like "Series," spatial data generating components like "Unit Z Vector," mathematical calculation components like "Power," spatial analysis components like "Area," and data structure processing components like "Flatten Tree." This comprehensive inclusion exposes Text2VP to a wide range of Grasshopper functions and datatypes.

Through the example output C# scripts, we demonstrate the general structure of C# coding syntax in the Grasshopper environment for generating workflows. This includes the syntax for creating and connecting Grasshopper components, positioning Grasshopper components on the canvas with designated spatial coordinates, directly assigning constant numeric values for inputs without connecting to another Grasshopper component, and controlling the display status of Grasshopper components, as a demonstration of the key points in the earlier instructions for customizing GPT (Section 2.2) with examples. Additionally, we draft extensive comments in the example output scripts to enhance Text2VP's understanding and learning.

These examples for few-shot learning are specifically prepared to present Text2VP with the conceptual framework of how the parametric modeling needs can be fulfilled using Grasshopper components from each function category. Additionally, they demonstrate the specific C# syntax required to automatically generate such workflows on the Grasshopper canvas.

## 3. Testing

This paper presents four testing samples for demonstrating 2D and 3D parametric modeling needs in the creasing complexity; they are simple yet representative for Text2VP to successfully produce the desired C# scripts, which can automatically generate the intended parametric modeling workflow on the Grasshopper canvas while utilizing most of the Grasshopper component function categories with proper coding syntax including those provided in the few-shot learning examples.

The four testing samples encompass various geometry modeling tasks, including 2D single curve object modeling to more complex building-like multiple 3D Boundary Representation (BREP) objects. Additionally, the samples include multiple data types connected and transformed between Grasshopper components through the visual programming workflow.

The testing samples are designed to gradually increase parametric modeling complexity. This progression is demonstrated by the transition from 2D to 3D modeling, the increasing number of interactable parameters, the increasing number of geometry objects, and the additional complexity of the logical structure of the visual program. This design allows us to test and evaluate Text2VP's ability to handle the increasing complexity of parametric modeling tasks.

### 3.1. Single-Object 2D Modeling

The first test is a simple single-object 2D modeling task: drawing and moving a circle with controls for radius and Z-axis location. The user prompt is as follows:

*I need a Grasshopper workflow to draw a circle on the XY plane and create a slider that controls the radius (2 to 20, default 20), and components with a slider to move the circle on the z-axis (-10 to 10, default 0).*

### 3.2. Single-Object 3D Modeling

The third test involves a straightforward single-object 3D modeling task that builds upon the previous test: creating a solid, top-flat cone with controls for top and bottom radii and height. The prompt is as follows:

*I need a Grasshopper workflow to draw a closed flattop cone. Include number sliders to control the bottom circle's radius (1 to 20, default 20, accuracy to the hundredth), the top circle's radius (1 to 20, default 10, accuracy to the thousandth), the height (5 to 10, default 7, accuracy to the tenth),*

### 3.3. Multi-Object 3D Modeling

The third test is a simple multi-object 3D modeling task developed from the second test: creating a multilayer conical tower-like 3D parametric model with controls for bottom radius, floor height, number of floors, and conical reduction factor. The prompt is as follows:

*I need a Grasshopper workflow to draw a closed round tower with each layer as a closed cylinder at a constant height. Each layer's base circle reduces the radius by a continuous reduction factor compared to the layer below. Include number sliders for the bottom circle radius (20 to 200, default 100, accuracy to the tenth), the height of each layer (1 to 20, default 10, accuracy to the hundredth), the total number of layers (1 to 20, default 10), and the reduction factor (0.1 to 1.0, default 0.75, accuracy to the thousandth).*

### 3.4. Recursive Multi-Object 3D Modeling

The fourth test is a more advanced, recursive multi-object 3D modeling challenge inspired by examples commonly found in parametric modeling tutorials and papers (Wang 2009; Woodbury 2010; Yan 2012): a multilayer tower-like 3D parametric model composed of recursively nested squares, with controls for bottom radius, floor height, number of floors, and rotation angles. The prompt is as follows:

*I need a Grasshopper workflow to draw a multilayer tower. Each layer is a closed, regular square prism with a constant height. The layers are based on a series of concentric squares, each rotated at constant angles and decreasing in size from the nearest outer square, with vertices always positioned at the edges of the nearest outer square. (Nested Squares) Include number sliders to control the bottom square's radius (20 to 200, default 100, accuracy to the tenth), layer height (1 to 20, default 10, accuracy to the hundredth), total layers (1 to 20, default 10), and control layer rotation (0 to 0.5, default 0.25, thousandth accuracy) for degrees between 0 and 90.*

## 4. Results and Discussion

In all tests, Text2VP consistently followed the prescribed process to analyze user prompts. In the end, it fulfilled its dedicated role of assisting in the generation of visual programming workflows for parametric modeling and responded accordingly.

However, our evaluation revealed that Text2VP does not always generate flawless C# scripts capable of automatically producing the user's intended workflows on the Grasshopper canvas in the initial attempt. From a prompt engineering perspective, refining the clarity and precision of the user prompt can improve the model's output. Additionally, in some cases, even without any modification to the user's prompt, providing the model with a second attempt resulted in complete scripts that successfully generated the intended visual programming workflows in Grasshopper without errors, thereby achieving the desired parametric modeling tasks.

### 4.1. Single-Object 2D Modeling

For this simple test case, Text2VP successfully generated C# scripts that produced the desired visual programming workflow on the Grasshopper canvas. The model fully met the parametric modeling objectives on the first attempt (Figures 3 and 4).

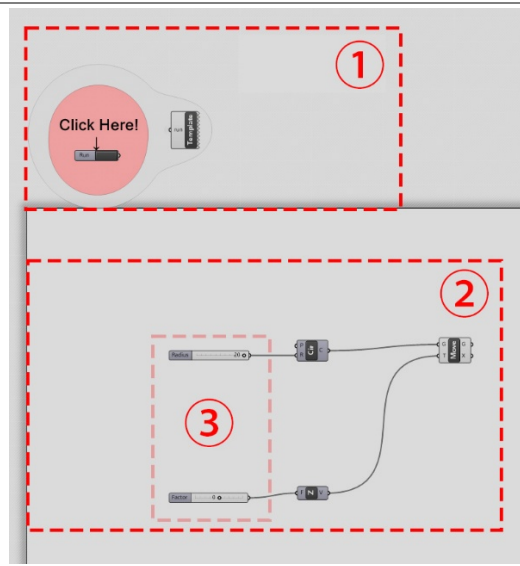

① Grasshopper C# Scripting Node ——"Workflow Generator"

② Grasshopper Visual Programming Workflow Automatically Generated by C# Script Text2VP Coded

③ Parametric Model Control Parameters

Figure 3. First Test: Grasshopper Canvas with C# Scripting Nodes and Visual Programming Workflow that Text2VP Automatically Generated.

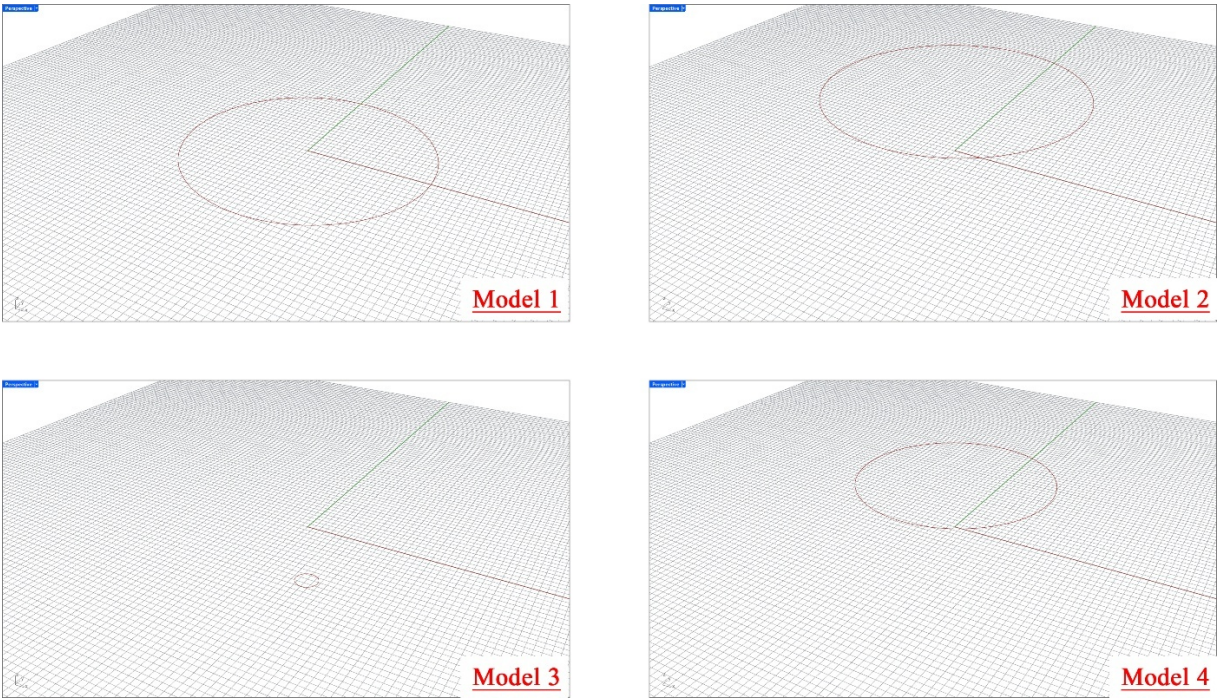

|  | Radius | Z-axis |
|---|---|---|
| Model 1 (default value) | 20 | 0 |
| Model 2 | 10 | 10 |
| Model 3 | 2 | -10 |
| Model 4 | 15 | 8 |

Figure 4. First Test: 4 Results Shown in Rhino by 4 Different Combinations of Parameters' Values in the Grasshopper Workflow that Text2VP Generated.

### 4.2. Single-Object 3D Modeling

In the second test, which involved a single-object 3D parametric modeling task, Text2VP demonstrated robust performance despite the increased complexity of transitioning from 2D to 3D modeling. The model generated complete, error-free C# scripts that successfully produced the intended visual programming workflow on the Grasshopper canvas in the first attempt, accurately fulfilling the parametric modeling objectives (Figures 5 and 6).

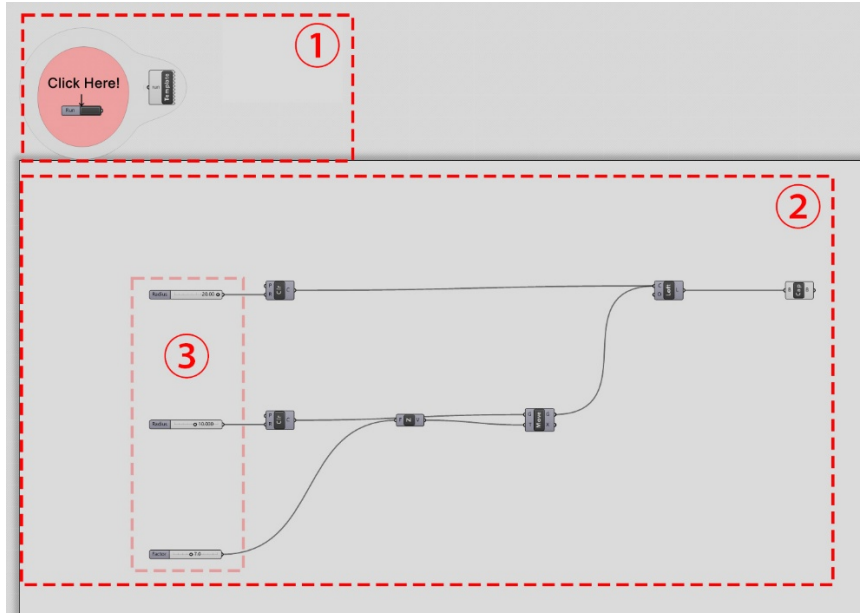

| ① | Grasshopper C# Scripting Node ——"Workflow Generator" |
| ② | Grasshopper Visual Programming Workflow Automatically Generated by C# Script Text2VP Coded |
| ③ | Parametric Model Control Parameters |

Figure 5.   Second Test: Grasshopper Canvas with C# Scripting Nodes and Visual Programming Workflow that Text2VP Automatically Generated.

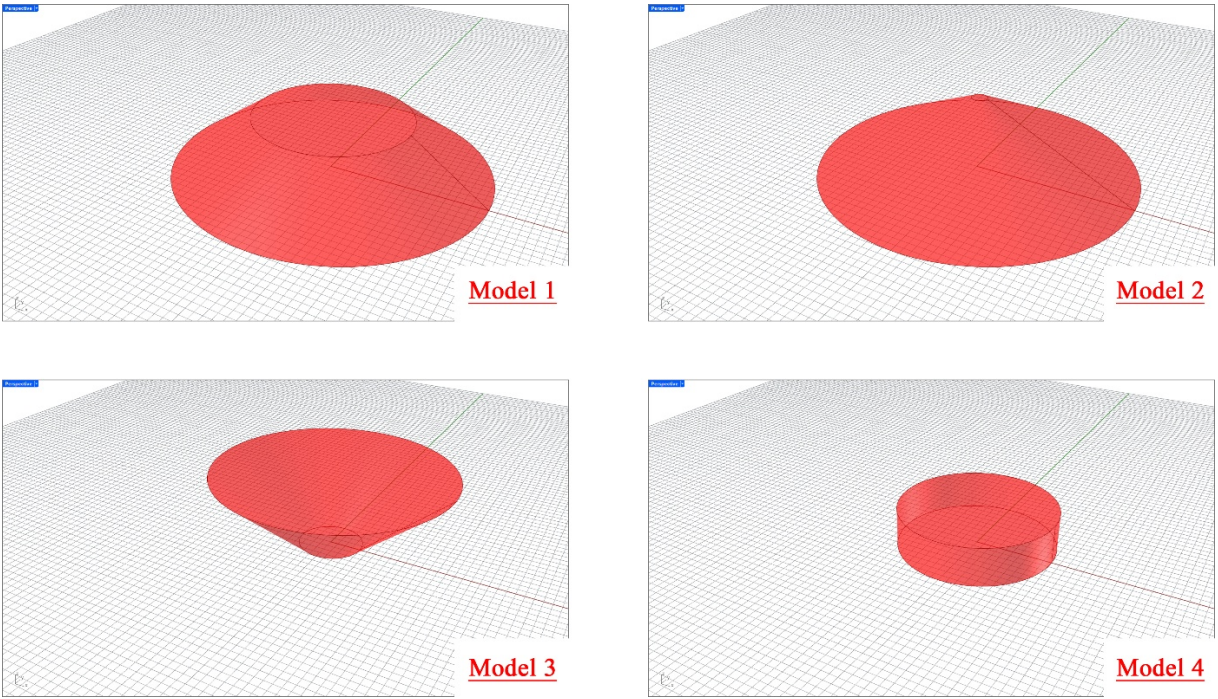

|  | Bottom Radius | Top Radium | Height |
|---|---|---|---|
| Model 1 (default value) | 20.00 | 10.000 | 7.0 |
| Model 2 | 20.00 | 1.000 | 9.5 |
| Model 3 | 4.00 | 15.000 | 9.5 |
| Model 4 | 10.00 | 10.000 | 5.0 |

Figure 6. Second Test: 4 Results Shown in Rhino by 4 Different Combinations of Parameters' Values in the Grasshopper Workflow that Text2VP Generated.

### 4.3. Multi-Object 3D Modeling

In the third test, which required generating a workflow for modeling multiple 3D objects, the complexity of the task further increased. On its first attempt, Text2VP did not produce scripts that generated an entirely error-free visual programming workflow on the Grasshopper canvas. A minor error was identified: the namespace for the Grasshopper "Extrude" component was incorrectly altered from "*SurfaceComponents.SurfaceComponents.Component_Extrude*" to "*SurfaceComponents.Component_Extrude*," a mistake attributed to AI hallucination. Despite this issue, the overall visual programming logic and structure of the scripts were correct. Notably, without any modification to the user prompt or the addition of further guidance, Text2VP succeeded on the second attempt, providing scripts that generated an error-free visual programming workflow on the Grasshopper canvas and accurately fulfilled the intended parametric modeling objectives (Figures 7 and 8).

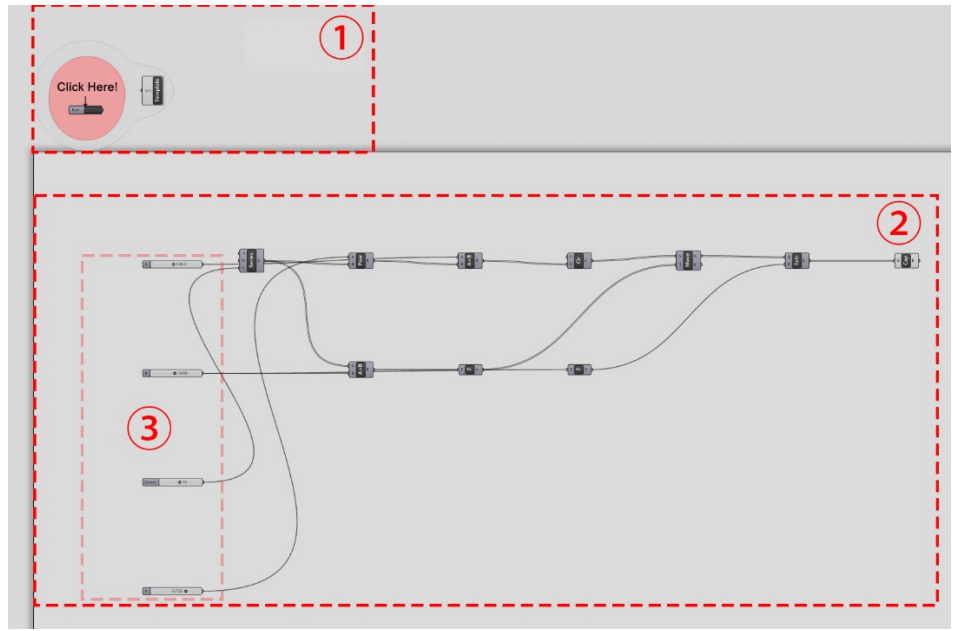

| | |
|---|---|
| ① | Grasshopper C# Scripting Node ——"Workflow Generator" |
| ② | Grasshopper Visual Programming Workflow Automatically Generated by C# Script Text2VP Coded |
| ③ | Parametric Model Control Parameters |

Figure 7.   Third Test: Grasshopper Canvas with C# Scripting Nodes and Visual Programming Workflow that Text2VP Automatically Generated.

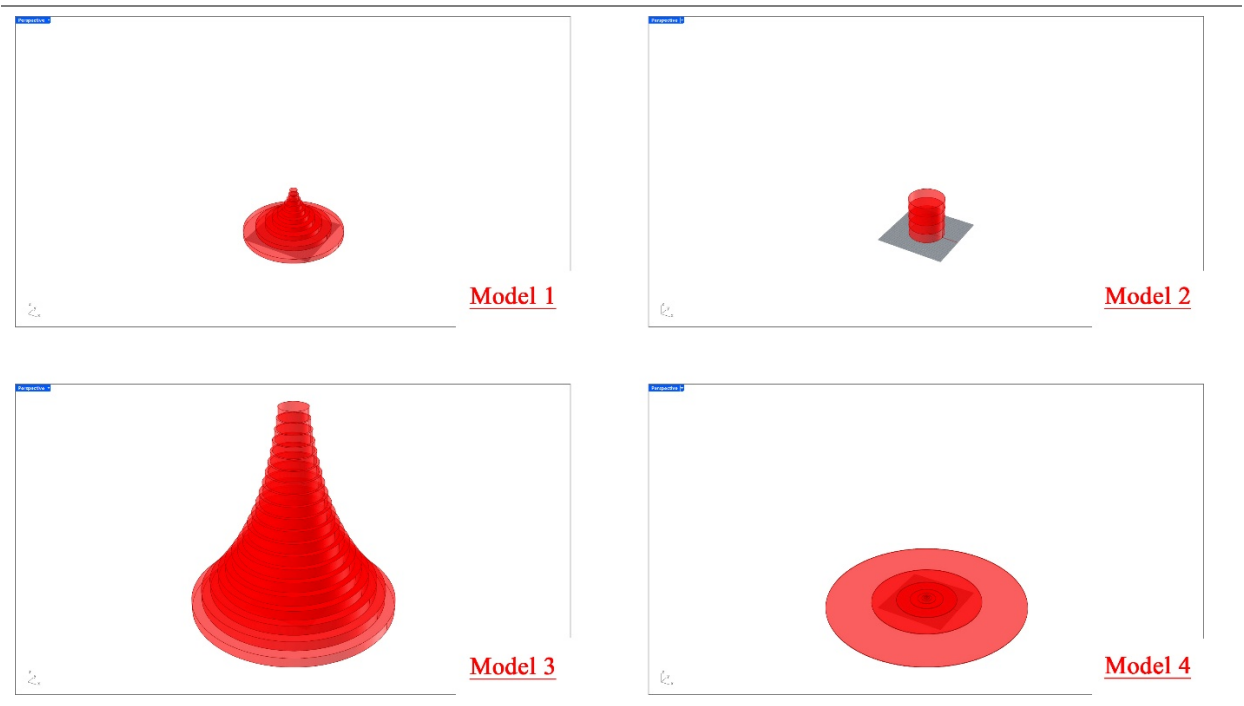

|  | Bottom Radius | Layer Height | Total Layers | Reduction Factor |
|---|---|---|---|---|
| Model1 (default value) | 100.0 | 10.00 | 10 | 0.750 |
| Model 2 | 35.7 | 20.00 | 4 | 1.000 |
| Model 3 | 200.0 | 20.00 | 20 | 0.900 |
| Model 4 | 200.0 | 1.00 | 10 | 0.550 |

Figure 8. Third Test: 4 Results Shown in Rhino by 4 Different Combinations of Parameters' Values in the Grasshopper Workflow that Text2VP Generated.

### 4.4. Recursive Multi-Object 3D Modeling

As the complexity of the task further increased, Text2VP's first attempt in the recursive multi-object 3D modeling task did not fully capture the intended modeling logic. Although the model produced a syntactically correct script and automatically generated a Grasshopper parametric modeling workflow on the canvas, the resulting model indicated that Text2VP struggled to accurately represent the geometric relationships involved in the recursive process—particularly the interplay between the squares' rotation and scaling as the "tower" rises. After revising the user prompt to provide a clearer explanation of the geometric recursion mechanism, the second attempt resulted in a visual programming algorithm that more closely fulfilled the parametric modeling requirements. However, a minor syntax error persisted, specifically involving the misuse of the connection syntax for the Grasshopper number slider component, which lacks output attributes and therefore requires a different connection syntax. Without any further modification to the user prompt or additional hints, Text2VP succeeded on the third attempt, generating scripts that produced an error-free visual programming workflow on the Grasshopper canvas and accurately fulfilled the parametric modeling objectives (Figures 9 and 10).

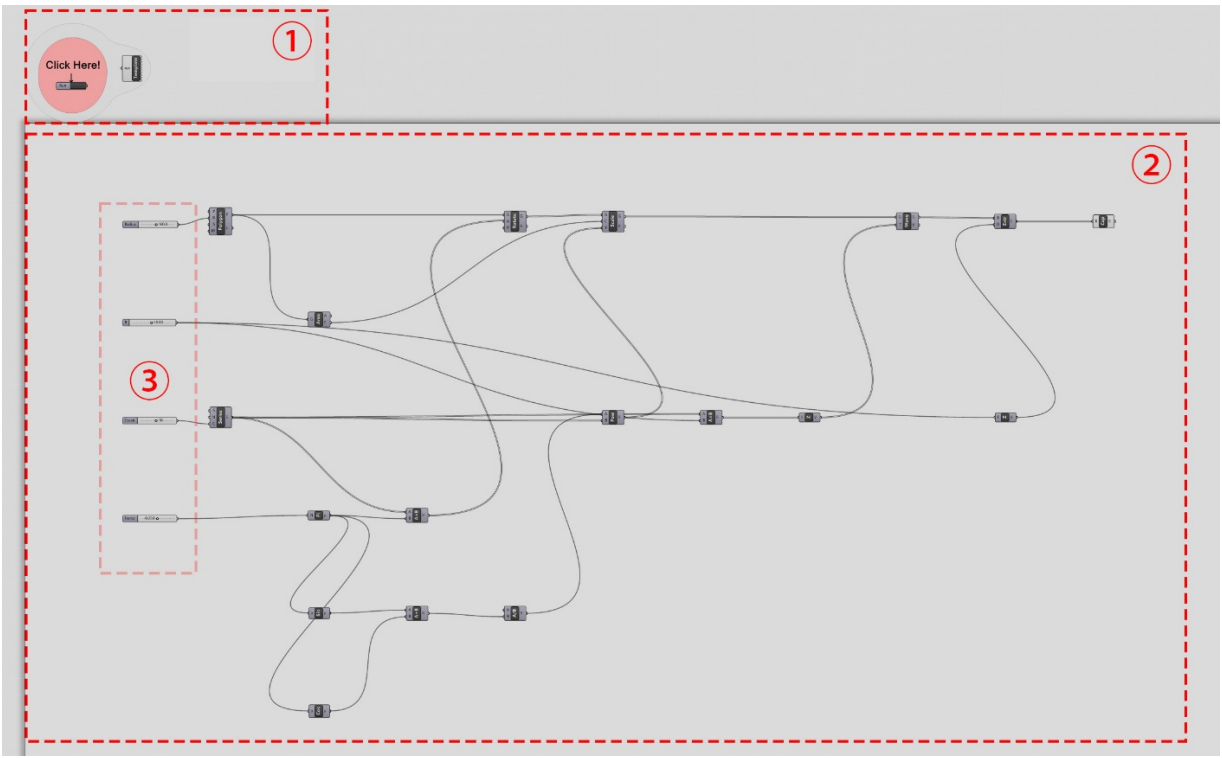

| ① | Grasshopper C# Scripting Node ——"Workflow Generator" |
|---|---|
| ② | Grasshopper Visual Programming Workflow Automatically Generated by C# Script Text2VP Coded |
| ③ | Parametric Model Control Parameters |

Figure 9.  Fourth Test: Grasshopper Canvas with C# Scripting Nodes and Visual Programming Workflow that Text2VP Automatically Generated.

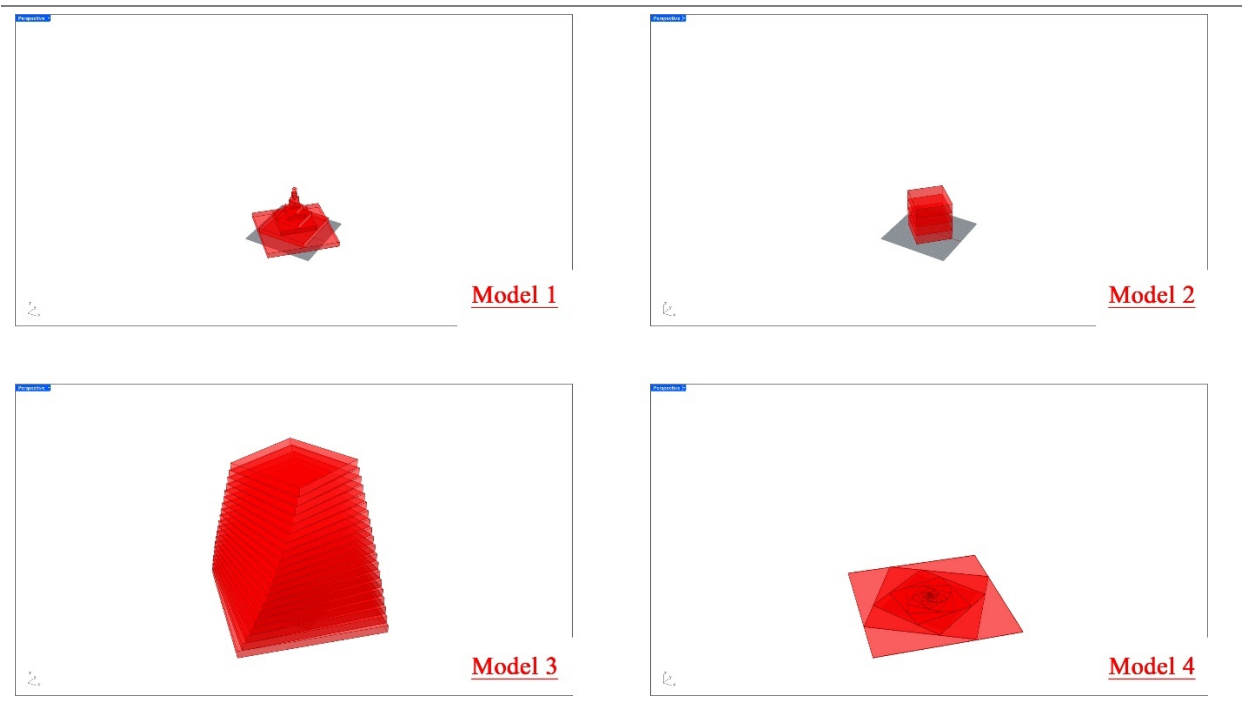

|  | Bottom Radius | Layer Height | Total Layers | Rotation Angle ($\pi$) |
|---|---|---|---|---|
| Model1 (default value) | 100.0 | 10.00 | 10 | 0.250 |
| Model 2 | 50.0 | 20.00 | 4 | 0.000 |
| Model 3 | 200.0 | 15.00 | 20 | 0.010 |
| Model 4 | 200.0 | 1.00 | 20 | 0.360 |

Figure 10. Fourth Test: 4 Results Shown in Rhino by 4 Different Combinations of Parameters' Values in the Grasshopper Workflow that Text2VP Generated.

Importantly, in all tests, once the Grasshopper workflows were generated automatically, users retained the ability to manually adjust parameter values—such as operating the sliders—to modify the resulting geometry in Rhino. Although some of the geometric models in the four sample tests may appear simple, the parametric modeling workflows produced by Text2VP are relatively sophisticated, as they incorporate interactive parameters, associated data types, and values that accurately reflect the user's design intentions. Notably, in the third and fourth test cases, the generated parametric models featured building-like geometries with considerable complexity, highlighting the significant potential of this approach for future applications in the AEC industry, as well as in other fields that visual programming and parametric modeling are utilized, such as mechanical, electrical, and system engineering design, game development, dataflow programming, workflow automation, education, etc.

## 5. Conclusions and Future Work

Despite this research producing limited testing samples, we can confidently conclude that this study is a success in customizing the Text2VP GPT model from the standard pre-trained GPT-4.1-2025-04-14 model (OpenAI 2025), using our specifically prepared knowledge base (Section 2.1) and instructions for customizing GPT (Section 2.2). The presented testing samples demonstrate that Text2VP adopts the knowledge and pattern from these materials, such as the functions of specific Grasshopper components and the syntax of C# coding in the Grasshopper environment. In all tests, Text2VP eventually produced fully working C# scripts that could be used in the Grasshopper environment to automatically generate visual programming workflows to accurately fulfill the user's parametric modeling intentions.

In summary, this research successfully introduces an innovative approach to applying generative AI to assist parametric modeling by automatically generating visual programming workflows that meet the user's design and modeling intent in prompts with natural human language (English). Rather than generating a fixed 3D model (e.g., a specific mesh model) from user prompts as the state-of-the-art (Tsalicoglou et al. 2023; Höllein et al. 2023; Guillard et al. 2021), this research advanced generative AI by automatically generating a parametric model with interactable parameters that control the parametric model, capable of generating multiple (e.g., millions) of models and design options for further simulation and optimization work. Additionally, for the first time, this research demonstrates the potential of generative AI in automating visual programming (which can be beyond Grasshopper, including Dynamo, LabView, Houdini, etc., visual programming languages), leveraging the widely recognized capabilities of generative AI in assisting coding tasks. This research also showcases how to teach the domain knowledge of visual programming to AI so that it can generate visual programs for parametric modeling or other applications.

While the research produced the first working prototype of AI-assisted generation of visual programs, the research is a preliminary exploration. The test results indicate that errors exist in Text2VP's responses, and these errors increase as the tasks become more complex. Additionally, during the testing process, there is no clear indication that Text2VP learns from its interactions with users regarding the errors in its responses, memorizes the corrections, and carries them over to tasks. However, this could be attributed to this research's very small testing sample size during the testing process; further testing may yield different conclusions. This highlights the need for more robust improvements and specific fine-tuning of generative AI, instead of few-shot learning as applied in this research, in future work, including more testing samples, to empower Text2VP to learn from its mistakes across tasks, thus reducing errors and eventually handling complex parametric modeling tasks accurately and effectively. The samples will include more classic examples in the textbook of parametric design: Elements of Parametric Design (Woodbury 2010) as benchmarks for testing and training Text2VP.

## Acknowledgments

To improve the clarity and grammatical accuracy of the textural content, we utilized the capabilities of ChatGPT (OpenAI version 4.1) [https://chat.openai.com/] for copy-editing purposes only.

## References

Asl, Mohammad Rahmani, Alexander Stoupine, Saied Zarrinmehr, and Wei Yan. 2015. "Optimo: A BIM-Based Multi-Objective Optimization Tool Utilizing Visual Programming for High Performance Building Design." *Proceedings of eCAADe 2015*, 673–82.

Brown, Tom, Benjamin Mann, Nick Ryder, Melanie Subbiah, Jared D Kaplan, Prafulla Dhariwal, Arvind Neelakantan, et al. 2020. "Language Models Are Few-Shot Learners." In *Advances in Neural Information Processing Systems*, 33:1877–1901. Curran Associates, Inc. https://proceedings.neurips.cc/paper/2020/hash/1457c0d6bfcb4967418bfb8ac142f64a-Abstract.html.

Castro Pena, M. Luz, Adrián Carballal, Nereida Rodríguez-Fernández, Iria Santos, and Juan Romero. 2021. "Artificial Intelligence Applied to Conceptual Design. A Review of Its Use in Architecture." *Automation in Construction* 124 (April): 103550. doi:10.1016/j.autcon.2021.103550.

Dong, Andy, and Alice M. Agogino. 1997. "Text Analysis for Constructing Design Representations." *Artificial Intelligence in Engineering* 11 (2): 65–75. doi:10.1016/S0954-1810(96)00036-2.

Ebert, Christof, and Panos Louridas. 2023. "Generative AI for Software Practitioners." *IEEE Software* 40 (4). IEEE: 30–38.

Gero, John S., and Thorsten Schnier. 1995. "Evolving Representations of Design Cases and Their Use in Creative Design." *Preprints Computational Models of Creative Design*. Key Centre of Design Computing, University of Sydney Sydney, Australia, 343–68.

Guillard, Benoit, Edoardo Remelli, Pierre Yvernay, and Pascal Fua. 2021. "Sketch2Mesh: Reconstructing and Editing 3D Shapes From Sketches." In , 13023–32. https://openaccess.thecvf.com/content/ICCV2021/html/Guillard_Sketch2Mesh_Reconstructing_and_Editing_3D_Shapes_From_Sketches_ICCV_2021_paper.html.

Höllein, Lukas, Ang Cao, Andrew Owens, Justin Johnson, and Matthias Nießner. 2023. "Text2room: Extracting Textured 3d Meshes from 2d Text-to-Image Models." In *Proceedings of the IEEE/CVF International Conference on Computer Vision*, 7909–20. http://openaccess.thecvf.com/content/ICCV2023/html/Hollein_Text2Room_Extracting_Textured_3D_Meshes_from_2D_Text-to-Image_Models_ICCV_2023_paper.html.

OpenAI. 2025. "GPT-4.1-2025-04-14." GPT. https://openai.com/index/gpt-4-1/.

"Retrieval - OpenAI API." 2025. Accessed May 15. https://platform.openai.com.

Robert McNeel & Associates (TLM, Inc.). 2023. "Rhinoceros." Microsoft Windows. English. Robert McNeel & Associates (TLM, Inc.). rhino3d.com.

Rudolph, Jürgen, Shannon Tan, and Samson Tan. 2023. "War of the Chatbots: Bard, Bing Chat, ChatGPT, Ernie and beyond. The New AI Gold Rush and Its Impact on Higher Education." *Journal of Applied Learning and Teaching* 6 (1). https://journals.sfu.ca/jalt/index.php/jalt/article/download/771/577/3333.

Russo, Daniel. 2024. "Navigating the Complexity of Generative AI Adoption in Software Engineering." *ACM Transactions on Software Engineering and Methodology*, March, 3652154. doi:10.1145/3652154.

Rutten, David. 2024. "Grasshopper." Windows. Robert McNeel and associates.

Saka, Abdullahi, Ridwan Taiwo, Nurudeen Saka, Babatunde Abiodun Salami, Saheed Ajayi, Kabiru Akande, and Hadi Kazemi. 2023. "GPT Models in Construction Industry: Opportunities, Limitations, and a Use Case Validation." *Developments in the Built Environment*. Elsevier, 100300.

Tagashira, K., N. Kawashima, and K. Yasuda. 2019. "Study on the Implementation of Visual Programming Language for Architectural Design Practices in Japan." *AIJ Journal of Technology and Design* 25 (60): 989–94.

Tsalicoglou, Christina, Fabian Manhardt, Alessio Tonioni, Michael Niemeyer, and Federico Tombari. 2023. "TextMesh: Generation of Realistic 3D Meshes From Text Prompts." arXiv. http://arxiv.org/abs/2304.12439.

Wang, Tsung-Hsien. 2009. "Design Patterns in GH." https://www.contrib.andrew.cmu.edu/org/tsunghsw-design/.

Woodbury, Robert (Robert Francis). 2010. *Elements of Parametric Design /*. Routledge,.

Wortmann, Thomas. 2018. "Efficient, Visual, and Interactive Architectural Design Optimization with Model-Based Methods." PhD Thesis, Singapore University of Technology and Design Singapore. https://www.researchgate.net/profile/Thomas-Wortmann/publication/327199280_Efficient_Visual_and_Interactive_Architectural_Design_Optimization_with_Model-based_Methods/links/5b7f697d299bf1d5a723c54d/Efficient-Visual-and-Interactive-Architectural-Design-Optimization-with-Model-based-Methods.pdf.

Yan, Wei. 2012. "Integrated BIM and Parametric Modeling: Course Samples with Multiple Methods and Multiple Phases." In *100th ACSA Annual Meeting Proceedings, Digital Aptitudes*, 198–207. https://www.acsa-arch.org/proceedings/Annual%20Meeting%20Proceedings/ACSA.AM.100/ACSA.AM.100.26.pdf.